\documentclass[10pt,letterpaper]{article}
\usepackage{opex3}
\usepackage{color}
\usepackage{amsmath,amssymb,graphicx}
\usepackage{epsfig}
\usepackage{epstopdf}

\newcommand{\be}{\begin{equation}}
\newcommand{\ee}{\end{equation}}
\newcommand{\bea}{\begin{eqnarray}}
\newcommand{\eea}{\end{eqnarray}}

\newcommand{\myeq}[1]{Eq.~(\ref{eq:#1})}
\newcommand{\myeqs}[1]{Eqs.~(\ref{eq:#1})}
\newcommand{\myeqss}[2]{Eqs.~(\ref{eq:#1}) and~(\ref{eq:#2})}

\begin{document}

\title{Paraxial ray optics cloaking}

\author{Joseph S. Choi$^{1,*}$ and John C. Howell$^{2}$}

\address{$^1$The Institute of Optics, University of Rochester, Rochester, New York 14627, USA\\
$^2$Department of Physics and Astronomy, University of Rochester, Rochester, New York 14627, USA}

\email{$^*$joseph.choi@rochester.edu} 

%





\begin{abstract}
Despite much interest and progress in optical spatial cloaking, a three-dimensional (3D), transmitting, continuously multidirectional cloak in the visible regime has not yet been demonstrated.
Here we experimentally demonstrate such a cloak using ray optics, albeit with some edge effects.  
Our device requires no new materials, uses isotropic off-the-shelf optics, scales easily to cloak arbitrarily large objects, and is as broadband as the choice of optical material, all of which have been challenges for current cloaking schemes. 
In addition, we provide a concise formalism that quantifies and produces perfect optical cloaks in the small-angle (`paraxial') limit.
\end{abstract}



\section{Introduction}
Optical spatial cloaking has captured the imagination of 
both the popular culture and scientific communities~\cite{Gbur-Cloak-2013}.
In particular, much scientific progress has been made recently in invisibility cloaking since the seminal works of Leonhardt~\cite{Leonhardt-2006} and Pendry, Schurig, and Smith~\cite{Pendry-2006}. They provided a theoretical framework to create a curved space for electromagnetic fields, by carefully constructing materials using coordinate transformations. This new field of research has been called `transformation optics'
~\cite{McCall-2013,Zhang-review-2012}. 
Experimental realization of these ideas has been difficult, due to the need for artificial electric and magnetic materials (called `metamaterials'), its narrow-band spectrum, infinite phase velocity (or negative index to compensate this), and anisotropy in the theory~\cite{Gbur-Cloak-2013}.
Nonetheless, inspired by transformation optics, there have been tremendous advances in cloaking. 
These include two-dimensional microwave cloaks~\cite{Schurig-2006,Landy-Smith-2013}, `carpet' cloaks that hide objects under a surface
~\cite{Li-Pendry-2008,Ergin-2010}, cloaking from scattering cancellation~\cite{Soric-Alu-Mantle-2013,Rainwater-Alu-3D-2012},
and even cloaking in time~\cite{Fridman-2012,Lukens-2013} or from seismic waves~\cite{Brule-2014}. 
A few groups have been able to cloak millimeter- to centimeter-sized objects as well, using birefringent materials~\cite{Zhang-2011,Chen-2011}.
In these developments to implement practical cloaks, Zhang has observed a shift from traditional metamaterials, to polymers, and to natural materials, suggesting that traditional optics can help to develop market-ready cloaking technologies~\cite{Zhang-review-2012}.  

In fact, to overcome the metamaterial requirements and to extend cloaking to a broadband, visible regime for large objects, researchers have recently looked to ray optics for cloaking~\cite{Chen-2013,Zhai-2013,Howell-Cloak-2014}. 
In these cloaks, the amplitude and direction of light fields are considered, as opposed to the full preservation of fields (amplitude and phase) of transformation optics. 
These designs have been able to cloak centimeter-~\cite{Chen-2013} to meter-sized~\cite{Howell-Cloak-2014} objects with commonly available optics. 
Yet, these schemes work only for unidirectionally incident light, or for discrete angles only, 
as they are not designed for continuously multidirectional cloaking, and they can have non-unity magnifications~\cite{Chen-2013,Zhai-2013,Howell-Cloak-2014}. 
For off-axis, non-zero angles, the background images show distortion and positional shifts.  This may not be clear when the background is close to the cloaking device~\cite{Chen-2013}, but becomes particularly evident when they are far apart. 
In addition, as can be seen in Fig. 1 of Reference~\cite{Howell-Cloak-2014}, 
rays that propagate through these cloaks, but go through the center at non-zero angles, can actually enter the cloaking region, effectively uncloaking the space.

Despite the advances in cloaking, a 3D multidirectional cloak has been elusive. As shown by Wolf and Habashy~\cite{Wolf-Cloak-1993} and Nachman~\cite{Nachman-1988}, no isotropic cloaking scheme can hide an object from \emph{all} viewing angles. Their works answered a question that stemmed from Devaney~\cite{Devaney-1978}, who elegantly showed how to mathematically construct potentials with zero scattering fields, hence invisible. However, Devaney's result was for a finite number of discrete directions, and not for a continuous range of angles.

\begin{figure*}[htbp]
\centerline{\includegraphics[width=1.0\columnwidth]{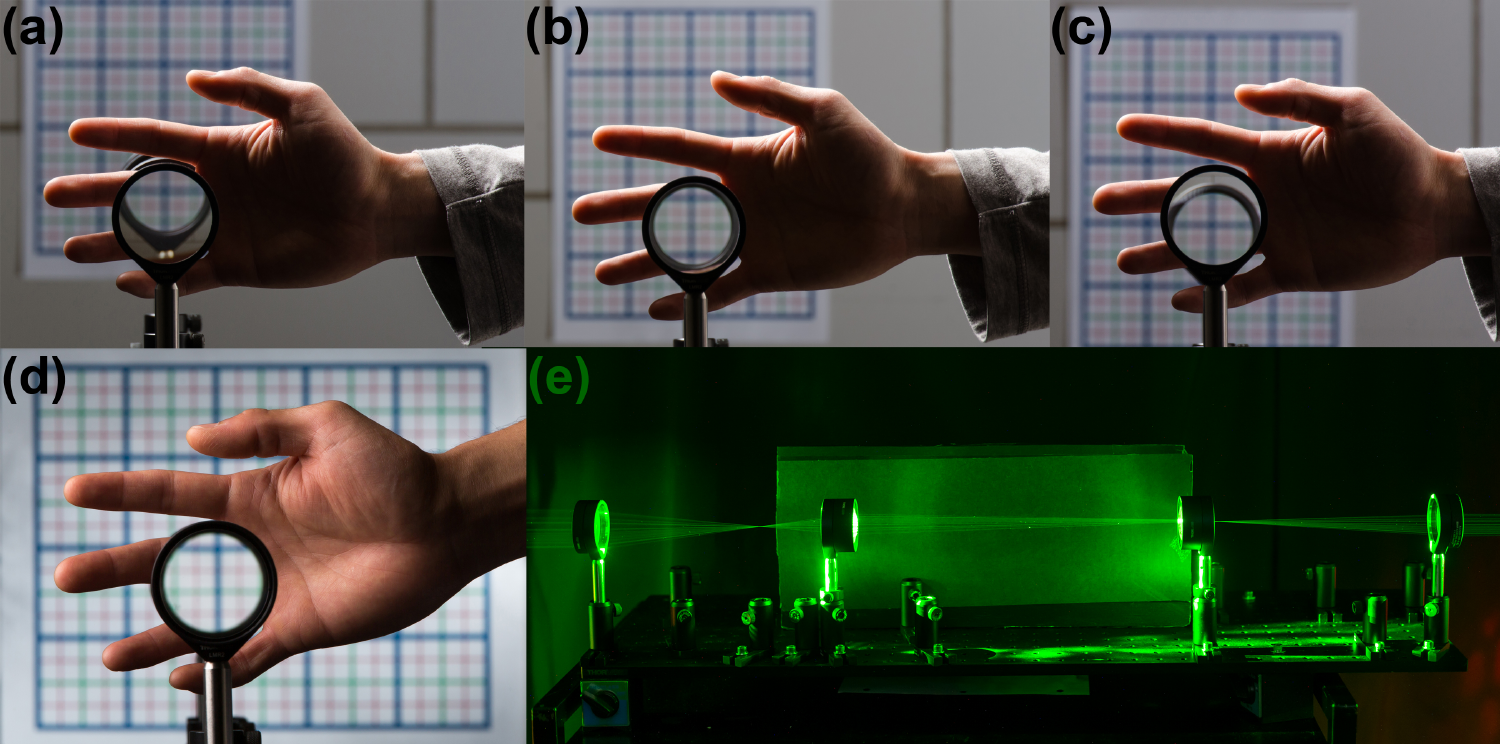}}
\caption{\label{fig:MultiClk2-4lens-UR}
\textbf{Example of a practical paraxial cloak.}
\textbf{(a)-(c)} A hand is cloaked for varying directions, while the background image is transmitted properly (See \textcolor{blue}{Media~1,2} for videos).
\textbf{(d)} On-axis view of the ray optics cloaking device.
\textbf{(e)} Setup using practical, easy to obtain optics, for demonstrating paraxial cloaking principles.
(Photos by J. Adam Fenster, videos by Matthew Mann / University of Rochester)}
\end{figure*}

In this article, we demonstrate a ray optics cloak that is designed for continuously multidirectional angles in 3D (See Fig.~\ref{fig:MultiClk2-4lens-UR}).  This is the first such device, to our knowledge, for transmitting rays in the visible regime.  It also uses off-the-shelf isotropic optics, scales easily to arbitrarily large sizes, has unity magnification, and is as broadband as the optical material used.  Thus, many of the difficulties encountered in invisibility cloaking schemes so far are solved, albeit with edge effects that are present. 
We also provide a concise and effective formalism, using ray optics, to describe all perfect optical cloaks in the small-angle (`paraxial') limit.
We apply our formalism to general optical systems up to four lenses, and show what systems can be considered `perfect' paraxial cloaks for rays. 

\section{Theoretical formalism}
To begin, we use a slightly different philosophy than transformation optics. Rather than starting with the bending or reshaping of the space for fields, we first consider replacing the cloaking space entirely.
If the cloaking device can be replaced in a simple manner, then engineering every field to move around the cloaked space may be unnecessary, or automatic.
This can be accomplished by considering the cloaking device as an optical system that images the background.
Our scope is limited to ray optics, so we do not attempt to preserve the complete phase of the fields.

\subsection{Defining a `perfect' cloak}
Let's first define a `perfect,' or `ideal,' cloaking device.
We take the broad definition of a ``cloak'' as something that ``hides'' an object or space, not necessarily as a garment to be worn or wrapped around an object. 
An obvious first requirement is that a `perfect' cloak must have a non-zero volume to hide an object.  
Second, such a cloak should act the same way as if it was not there. 
This is equivalent to the device being replaced completely by the ambient medium.
These two conditions then are sufficient and necessary for defining a \emph{`perfect cloak'} in general. 
For such, both the cloaked object and the cloaking device are invisible~\cite{Leonhardt-2006,Schurig-2006}.

\begin{figure}[htbp]
 \begin{centering}
   \includegraphics[scale=1]{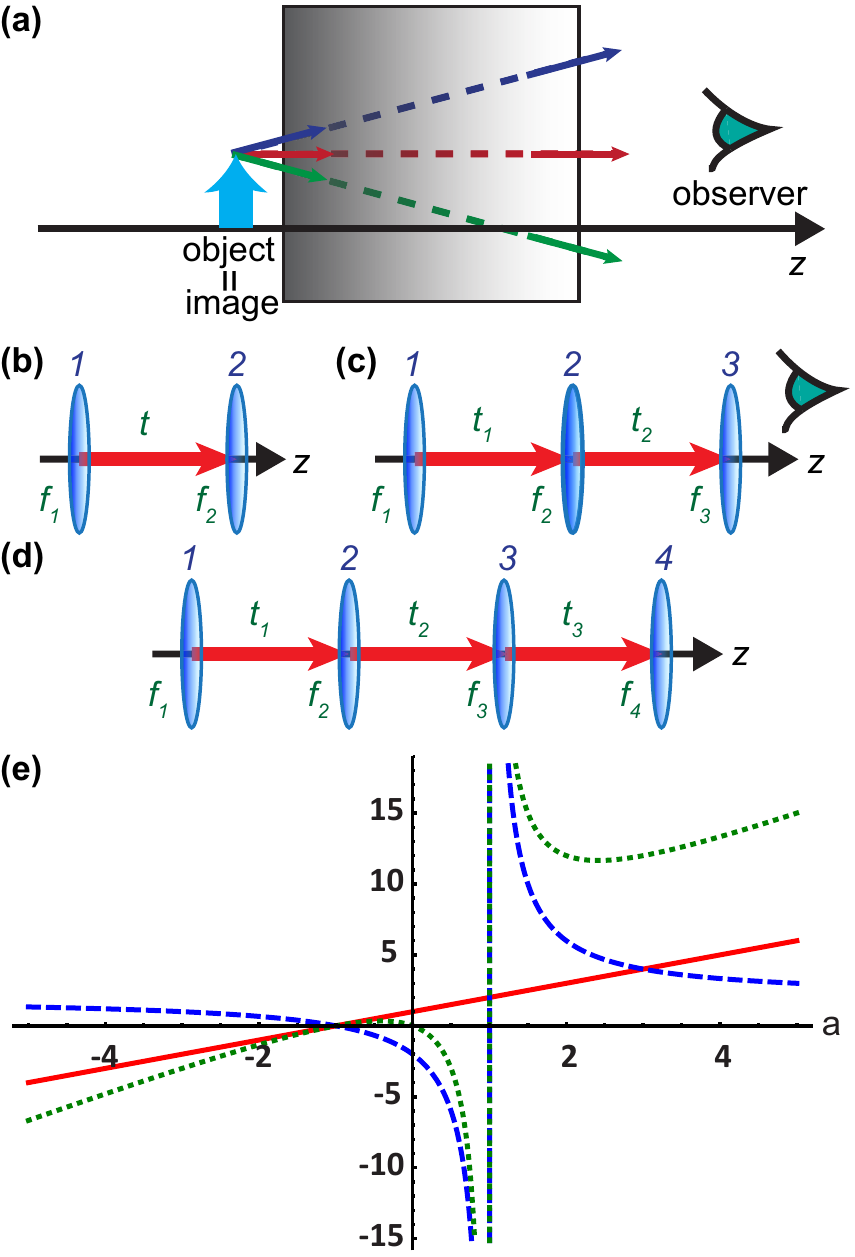}
\caption{\label{fig:MultiClk2-clks}
\textbf{Investigating `perfect' paraxial cloaking with rays.}
\textbf{(a)} A `perfect' ray optics cloaking box. Rays exit the box as if the box was filled with the surrounding medium. Non-zero volume inside hides an object. Angles do not change, but the positions shift proportionally to the ray angles and box length. The image seen by the observer should match the object exactly.
\textbf{(b)-(d)} Diagrams for a two lens \textbf{(b)}, 
three lens \textbf{(c)}, or
four lens \textbf{(d)} system.
$f$'s are the focal lengths, $t$'s are the distances between the elements.
\textbf{(e)} All possible four lens, symmetric, perfect paraxial cloaks for rays.
Plot of $t_1/f_2$ (solid), $t_2/f_2$ (dashed), and $L/f_2$ (dotted) as a function of $\textsf{a} \equiv f_1/f_2$. 
Assumed symmetric left and right halves ($f_1 = f_4$, $f_2=f_3$, and $t_1=t_3$). 
$L$ is the total length of the system.
The physical feasibility and presence of a non-empty cloaking region must be checked separately.
}
  \end{centering}
\end{figure}

We now discuss what a perfect cloak would do to light rays in the ray optics picture.  
According to our definition, such a device should behave as if its space was replaced by the surrounding medium.
Then, the ray angles exiting the device would not change, but the ray positions do shift (See Fig.~\ref{fig:MultiClk2-clks}(a)).
The image of an object behind the device, as seen by an observer, would be identical to the object itself.
This implies that the image location, size, shape, and color should be exactly that of the actual object.
A perfect ray optics cloak would generate images with unity magnification, zero transverse and longitudinal shifts, and no aberrations, i.e., \emph{no} changes, compared to the actual object, for \emph{all} ray positions and directions.

\subsection{Quantifying a perfect paraxial cloak}
So far, our definition of a `perfect' cloak was applicable generally. 
We will now develop a formalism using geometric optics, to quantify this definition in the paraxial approximation. 
To first-order approximation, called the ``paraxial approximation,'' light rays are assumed to deviate minimally from the center axis of the system. Hence, it is a small-angle approximation.
In this regime, also known as ``Gaussian optics''~\cite{Born-Wolf-2010}, propagation of light rays through an optical system can be described by `ABCD' matrices (see Fig.~\ref{fig:ABCD-intro} in \ref{app:ABCD-intro})~\cite{Siegman-book-1986,Handbook-Opt-v1-2010}.
Because a perfect cloaking device simply replicates the ambient medium throughout its volume, its ABCD matrix is just a `translation matrix': 
\begin{equation}
\begin{bmatrix}
A & B \\
C & D
\end{bmatrix}_{\text{perfect cloak}}
=
\begin{bmatrix}
1 & L/n \\
0 & 1
\end{bmatrix}
.
\label{eq:cloak-metric}
\end{equation}
$L$ is the length of the cloaking system, and $n$ is the index of refraction of the surrounding medium (For a nonuniform ambient medium, see \myeq{cloak-metric-nonuniform} in \ref{app:ABCD-cloak}).

Equation~(\ref{eq:cloak-metric}) is at the heart of our paper.
Since any paraxial system can be written with ABCD matrices, if an optical system meets \myeq{cloak-metric} and has a cloaking region, then it \emph{is} a perfect paraxial cloak.
Any `perfect' cloak should also be perfect in first-order, so \myeq{cloak-metric} is a necessary condition for all such cloaks. 
Note that this does not violate the findings by Wolf and Habashy~\cite{Wolf-Cloak-1993} and Nachman~\cite{Nachman-1988} since this is a paraxial approximation, and hence does not work for large angles. However, we show this to be a surprisingly effective condition, despite its simplicity. 

Because ABCD matrices have a determinant of $1$, \myeq{cloak-metric} gives only three conditions to be satisfied:
$B=L/n$, $C=0$, and
\emph{either} $A=1$ \emph{or} $D=1$.
Note that a perfect cloaking system is ``afocal'' ($C=0$), meaning the optical system has no net focusing power. So an object at infinity will be imaged to infinity. This is helpful for the design process, since an afocal condition can be easily checked.

It is worth distinguishing a `perfect' paraxial cloak (\myeq{cloak-metric}) from a `perfect' cloak. In the paraxial regime, ray optics is used, there are no aberrations by definition, and sags or edges of optics are ignored because of the small-angle limit~\cite{Handbook-Opt-v1-2010,FG-GO-2004}.
However, real optics are not paraxial only, so a perfect paraxial cloak will have aberrations (non-ideal images), and unwanted rays may be visible near the edges (what we term as `edge effects'). 
On the other hand, a `perfect' cloak would hide an object entirely from the full field (amplitude and phase), so no changes to the field can be observed, including any aberrations.
These distinctions do go away for small-angle, nearly on-axis rays, and large optics.

\subsection{Designing a perfect paraxial cloak with rays}
It is not obvious that an optical system can satisfy \myeq{cloak-metric}, despite containing a cloaking region. The discussion and conditions for a `perfect cloak' may have little meaning unless a physical solution actually does exist. 
We will now carefully build general optical systems, to see whether a perfect paraxial cloak can be designed with rays.
We attempt to find the simplest nontrivial solution, so we will only consider rotationally symmetric systems with thin lenses, and in free space with $n=1$.

The ABCD matrix for one thin lens is given by
\be
\begin{bmatrix}
A & B \\
C & D
\end{bmatrix}_{\text{thin lens}}
=
\begin{bmatrix}
1 & 0 \\
-1/f & 1
\end{bmatrix}
,
\label{eq:ABCD-lens}
\ee
where $f$ is the focal length of the lens. We can easily see that \myeq{cloak-metric} will only be satisfied if $f=\pm \infty$, i.e. the lens has no optical power (it's flat). This has no cloaking region and no optical effect.


For the following, we will use $f$'s to denote the focal lengths, and $t$'s to denote the distances between the optical elements (Fig.~\ref{fig:MultiClk2-clks}(b)-(d)). 
The ABCD matrix for a two lens system 
(Fig.~\ref{fig:MultiClk2-clks}(b)) 
is
\bea
\begin{bmatrix}
 1 & 0 \\
 -1/f_2 & 1 \\
\end{bmatrix}
.
\begin{bmatrix}
 1 & t \\
 0 & 1 \\
\end{bmatrix}
.
\begin{bmatrix}
 1 & 0 \\
 -1/f_1 & 1 \\
\end{bmatrix}
=\nonumber
\\
\begin{bmatrix}
 1-t/f_1 & t \\
 -(f_1+f_2-t)/(f_1 f_2) & 1-t/f2 \\
\end{bmatrix}
.
\label{eq:ABCD-2lens}
\eea
Equation~(\ref{eq:cloak-metric}) will only be satisfied if $f_1=f_2=\pm \infty$. This is a system that is essentially made of empty space only, quite literally, again with no cloaking region nor optical effect.

A three lens system 
(Fig.~\ref{fig:MultiClk2-clks}(c)) 
has the following ABCD matrix:
\bea
\begin{bmatrix}
 1 & 0 \\
 -1/f_3 & 1 \\
\end{bmatrix}
\!.\!
\begin{bmatrix}
 1 & t_2 \\
 0 & 1 \\
\end{bmatrix}
\!.\!
\begin{bmatrix}
 1 & 0 \\
 -1/f_2 & 1 \\
\end{bmatrix}
\!.\!
\begin{bmatrix}
 1 & t_1 \\
 0 & 1 \\
\end{bmatrix}
\!.\!
\begin{bmatrix}
 1 & 0 \\
 -1/f_1 & 1 \\
\end{bmatrix}
\!\!.
\label{eq:ABCD-3lens}
\eea
We can solve for $f_2$ by setting $C=0$:
\be
f_2 = 
-\frac{(f_1 - t_1) (f_3 - t_2)}{f_1 + f_3 - t_1 - t_2}
.
\label{eq:3lens-f2C}
\ee
Using \myeq{3lens-f2C}, the ABCD matrix becomes
\bea
\begin{bmatrix}
\frac{f_3 (f_1 - t_1)}{f_1 (f_3 - t_2)}
&
t_1 + t_2 + t_1  t_2\frac{(f_1 + f_3 - t_1 - t_2)}{(f_1 - t_1) (f_3 - t_2)}
\\
0
&
\frac{f_1 (f_3 - t_2)}{f_3 (f_1 - t_1)}
\\
\end{bmatrix}
.
\label{eq:ABCD-3lens-f2C}
\eea
Requiring $B=t_1+t_2$ then gives
\be
t_1  t_2\frac{(f_1 + f_3 - t_1 - t_2)}{(f_1 - t_1) (f_3 - t_2)}
=0
.
\label{eq:3lens-f2B}
\ee
However, this is only true if $t_1=0$, or $t_2=0$, or if $(f_1 + f_3 - t_1 - t_2)=0$. 
The first two cases give the two lens system which we already showed cannot be a perfect cloak. 
The last case makes $f_2 \rightarrow \infty$, which also turns this system into a two lens system.

Although a three lens system cannot be a perfect cloak, it can asymptotically approach a paraxial one. 
For simplicity, let us consider the case with symmetric halves ($f_1=f_3, t_1=t_2$).
Then, \myeq{3lens-f2C} becomes
\be
f_2 = 
(t_1 - f_1)/2
,
\label{eq:3lens-f2C-symm}
\ee
and \myeq{3lens-f2B} becomes
\be
2 t_1^2 / (f_1 - t_1)
=0
.
\label{eq:3lens-f2B-symm}
\ee
So for $f_1 \gg t_1$, both \myeqss{3lens-f2C-symm}{3lens-f2B-symm} can be satisfied in this limit.
We will demonstrate the practicality of this case later.

Lastly, we consider a four lens system 
(Fig.~\ref{fig:MultiClk2-clks}(d)). 
We desire to undo any changes that the first half of our system makes, as a possible strategy to make the system behave as if absent.
This can be done by making the second half symmetric to the first half ($f_1=f_4$, $f_2=f_3$, $t_1=t_3$). 
We then require $A=1$ and $C=0$ for such a four lens ABCD matrix. Both conditions are conveniently satisfied by
\be
t_1 = f_1 + f_2
.
\label{eq:4lens-t1-symm}
\ee
With \myeq{4lens-t1-symm}, the ABCD matrix becomes
\bea
\begin{bmatrix}
1
&
f_1 (-2 t_1^2 + f_1 (2 t_1 + t_2))/(f_1 - t_1)^2

\\
0
&
1
\\
\end{bmatrix}
.
\label{eq:ABCD-3lens-f2C}
\eea
We can now set 
$B=(2 t_1 + t_2)$, 
and solve for $t_2$:
\be
t_2 = 2 f_2 (f_1 + f_2) / (f_1 - f_2)
.
\label{eq:4lens-t2-symm}
\ee

We have finally found an exact solution for \myeq{cloak-metric}.
At least four lenses are required for a perfect paraxial cloak, for a rotationally symmetric lens-only system.
With \myeqss{4lens-t1-symm}{4lens-t2-symm}, the total length is
\be
L = 2 t_1 + t_2 
= 2 f_1 (f_1 + f_2) / (f_1 - f_2)
.
\label{eq:4lens-L-symm}
\ee

We plotted $t_1, t_2$, and $L$ (\myeqs{4lens-t1-symm}, (\ref{eq:4lens-t2-symm}), and (\ref{eq:4lens-L-symm})) in 
Fig.~\ref{fig:MultiClk2-clks}(e).  
The figure describes all possible symmetric, four lens, perfect paraxial cloaks using rays. 
For now, we note only a few points of interest. 
We can see that when $f_1 \rightarrow -f_2$, the system approaches a one lens system ($t_1=t_2=L=0$). 
The two extrema for $L$, seen in 
Fig.~\ref{fig:MultiClk2-clks}(e), 
occur when
\be
f_1= 
(1\pm\sqrt{2}) f_2
.
\label{eq:4lens-L-extrema}
\ee
Although these solutions satisfy \myeq{cloak-metric} mathematically, checks must be made to ensure they contain a finite cloaking region and are physically feasible. 

\section{Experiments and simulations}
We now demonstrate two cloaks using the paraxial cloaking formalism that was developed. Recall that our formalism was for rotationally symmetric systems.  Hence, to show continuous multidirectionality in 3D, we need only demonstrate the cloaking effect by varying angles along one transverse direction.  Viewing angles along all other transverse directions, spanning a 3D range of incident angles, are the same due to rotational symmetry.

\subsection{Three lens cloak}
\begin{figure}[htbp]
\begin{centering}
\includegraphics[scale=1.0, angle=0, trim= 0 0 0 0]{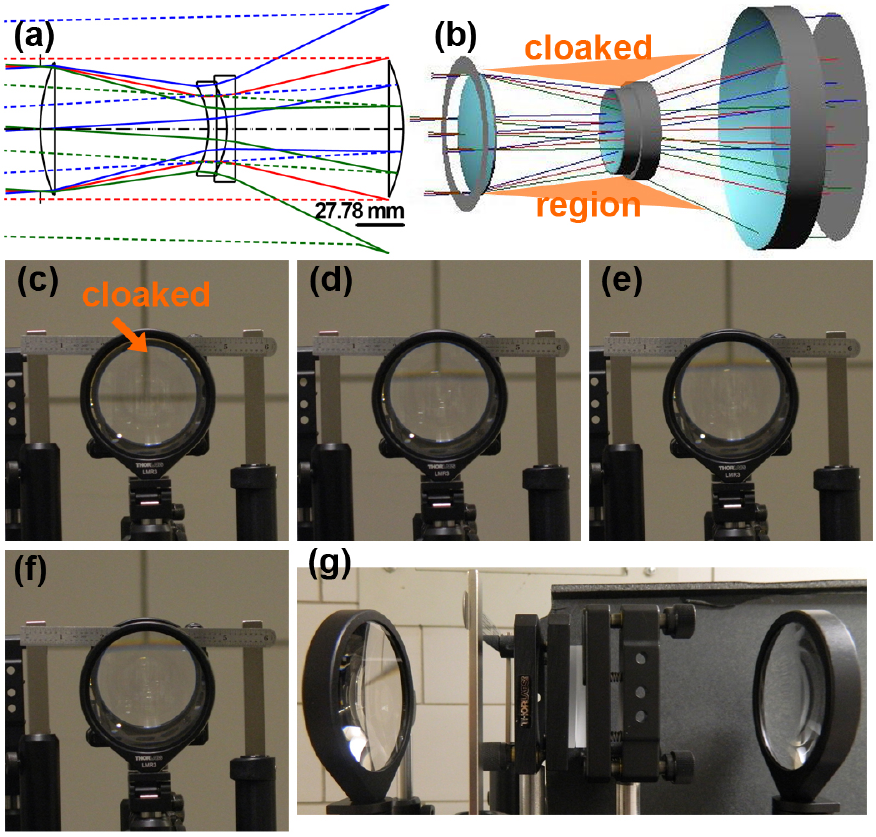}
\caption{\label{fig:MultiClk2-3lens}
\textbf{A symmetric three lens cloak.}
Two diverging lenses are combined into one diverging lens, and placed in the center of two converging lenses.
\textbf{(a)} 
Simulation in CODE V.
Entrance pupil is 75 mm, and field-of-view is $-3.5^\circ$ to $3.5^\circ$. 
Object is placed at infinity.
Ray bundles propagate from left to right, through the lenses, then are traced back to the first lens. This allows comparison of the image (dashed) rays, as seen by an observer on the right, with the original (solid) rays. We see that the angles are similar, and the transverse shifts are not large. 
\textbf{(b)} 3D rendering of \textbf{(a)}. The cloaking region is a 3D triangular-ring between the first and last lenses (shaded area).
\textbf{(c-g)}
Experimental demonstration of the three lens cloak.
The lines seen through the lenses match those on the background wall. The inner portion of the ruler is cloaked.
Images at various camera-viewing angles:
\textbf{(c)} On-axis ($0^\circ$),
\textbf{(d)} $0.55^\circ$,
\textbf{(e)} $0.83^\circ$,
\textbf{(f)} $1.11^\circ$.
\textbf{(g)} Side profile of experimental setup.
}
\end{centering}
\end{figure}

The cloaking region for our lens designs depend on what incident angles, or ``field-of-view,'' are allowed. To view the size of the cloaking space, we provide ray-trace simulations using CODE V.
We first simulate a three lens cloak with symmetric left and right halves (Fig.~\ref{fig:MultiClk2-3lens}(a),(b)). Recall that this system approaches a perfect paraxial cloak as its length goes to zero. 
Details of both the three lens and four lens setups are provided in \ref{app:methods}.
The size of the ray bundle entering the system (``entrance pupil'') was set to the first lens diameter in our experimental setup (75 mm).
The field-of-view is $-3.5^\circ$ to $3.5^\circ$. 
The cloaking region is between the lenses and is the ring-shaped region where no rays pass.
Fig.~\ref{fig:MultiClk2-3lens}(a) compares the final image rays to the original rays near the first lens. 
For a perfect cloak, these rays would overlap exactly. We can see that the angles are similar, and the transverse shifts are not too large.

We used plano-convex and plano-concave lenses for the three lens cloak. 
For the experimental demonstration (Fig.~\ref{fig:MultiClk2-3lens}(c)-(g)), the object (wall) was approximately 2 m from the closest lens in the back. The camera was 
5.3 m away from the front lens, but optically zoomed in by 21x (the maximum magnification of the camera). The images were taken from on-axis ($0^\circ$), $0.55^\circ$, $0.83^\circ$, and $1.11^\circ$, by increasing the height of the camera.
A ruler was placed near the center diverging lenses. 
We can see that the middle of the ruler is cloaked. Also, the lines of the wall match the background wall, as expected for a good cloak. 


\subsection{`Perfect' paraxial four lens cloak}
\begin{figure}[htbp]
\begin{centering}
\includegraphics[scale=1, angle=0, trim= 00 00 00 00]{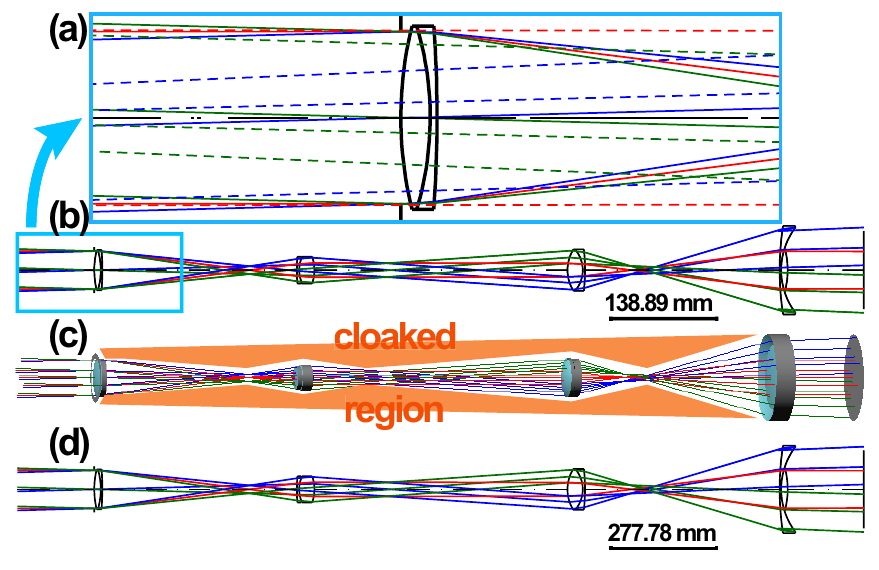}
\caption{\label{fig:MultiClk2-4lens-sim}
\textbf{CODE V simulation of a symmetric, perfect paraxial cloak, with four lenses using rays.}
Four achromatic doublets are placed with separations determined from \myeq{cloak-metric}.
Entrance pupil is 50 mm, with $-1.5^\circ$ to $1.5^\circ$ field-of-view. 
Simulations are shown with \emph{no} separate optimization. Object is placed at infinity.
\textbf{(a)} Zoomed-in region of \textbf{(b)} with image rays (dashed; traced back to the first lens) added to compare with the original rays (solid). We see that the angles are nearly identical, and the transverse shifts are small. 
\textbf{(b)} Full simulation using off-the-shelf optics.
\textbf{(c)} 3D rendering. The cloaking region (shaded) is a cylindrical region between the first and last lenses.
\textbf{(d)} Scaling of \textbf{(b)} by a factor of 2. The cloaking size is doubled by doubling the optical curvatures, lengths and entrance pupil. Only the length scales distinguish \textbf{(d)} from \textbf{(b)}.}
\end{centering}
\end{figure}

We now simulate a four lens `perfect' paraxial cloak for our experimental setup, that has symmetric left and right halves.
Real lens systems produce aberrations that can blur and distort the observed image. So we used `achromatic doublets' that combine two lenses as one, to correct for chromatic (color) and other aberrations.
We corrected \myeqss{4lens-t1-symm}{4lens-t2-symm} to include the lens thicknesses, and calculated $t_1, t_2$, and $t_3$ ($t_1=t_3$). The simulations in Fig.~\ref{fig:MultiClk2-4lens-sim} use only these calculated paraxial values without any additional optimization. The cloaking region is an elongated cylinder between the lenses where the rays do not pass.

Finally, we demonstrate the ease of scalability for our designs, to fit any cloaking size. We only need to scale all radii of curvature, lengths, and entrance pupil by the same factor. In Fig.~\ref{fig:MultiClk2-4lens-sim}(d), we simply doubled all of these parameters to obtain double the cloaking space.

\begin{figure}[htbp]
\begin{centering}
\includegraphics[scale=1.0, angle=0, trim= 00 00 00 00]{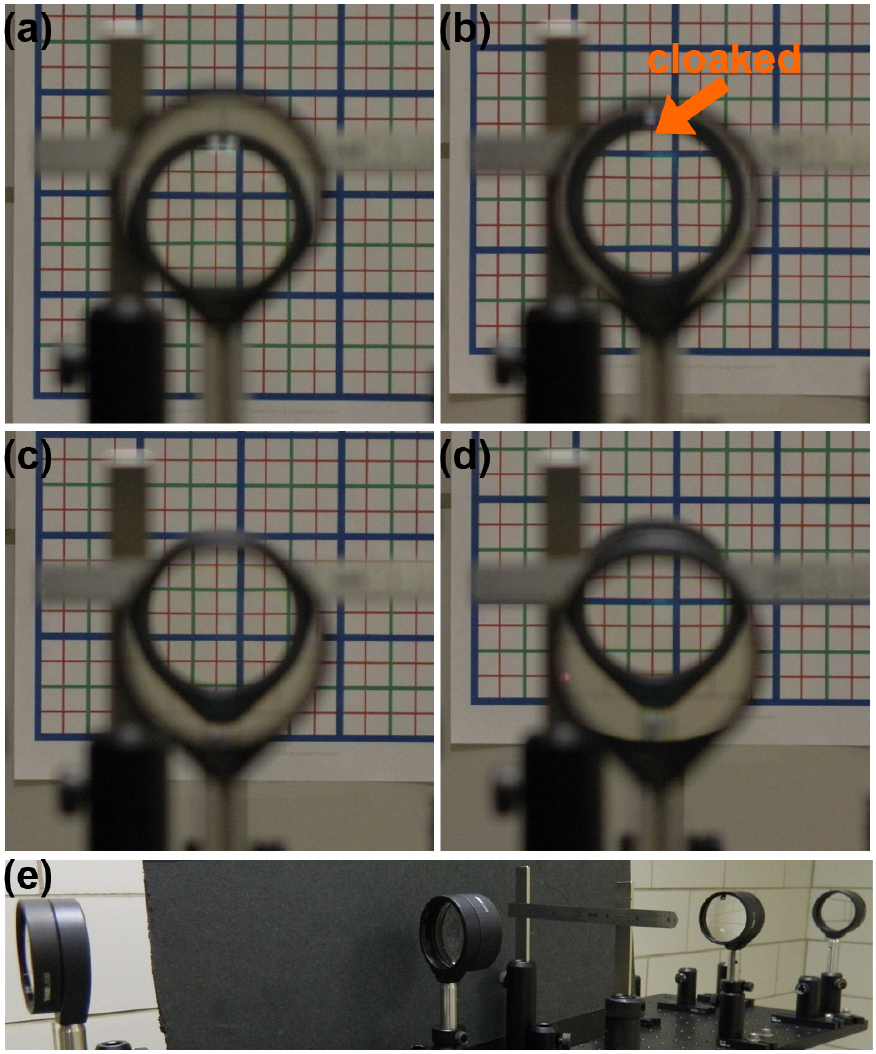}
\caption{\label{fig:MultiClk2-4lens-exp}
\textbf{Experimental demonstration of a `perfect' paraxial cloak with four lenses.}
Camera was focused on the wall. The grids on the wall can be seen clearly, and match the background for all colors and viewing angles. The middle of the ruler is cloaked inside the lens system for all angles shown.
Images at various camera-viewing angles:
\textbf{(a)} $-0.65^\circ$, 
\textbf{(b)} on-axis ($0^\circ$),
\textbf{(c)} $0.47^\circ$,
\textbf{(d)} $0.95^\circ$.
\textbf{(e)} Side profile of experimental setup.
See \textcolor{blue}{Media~3,4} for videos.
(Videos by Matthew Mann / University of Rochester)
}
\end{centering}
\end{figure}

In constructing our four lens cloak, we used achromatic doublets to reduce the aberrations of the images. Photographs of this paraxial cloak are shown in Fig.~\ref{fig:MultiClk2-4lens-exp}.
The grids on the wall were 1.9 m from the closest lens to the back. 
The camera was 3.1 m away from the front lens, but optically zoomed in by 21x. 
The images were taken from $-0.65^\circ$, on-axis ($0^\circ$), at $0.47^\circ$, and $0.95^\circ$ viewing angles, by changing the height of the camera.
A ruler was placed behind the second doublet from the front. 
The middle of the ruler is cloaked near the center-axis of the device. 
In particular, the grids on the wall are clear for all colors, have minimal distortion, and match the sizes and shifts of the background grids for all the angles, demonstrating the quality of this multidirectional cloak. 

\section{Limitations}
It is important to state the limitations of our current designs so that future work can improve on these.
The broadband capability of our designs is only limited by the coating and material used. This can be quite large, spanning the whole visible spectrum and beyond. However, to maintain clear images for an extended spectrum, well-designed achromats and coatings, combined with other lens design techniques are necessary to correct for aberrations.  Here, a study of a particular ``NonLens'' problem, which satisfies \myeq{cloak-metric} but without requiring the presence of a cloaking volume, may be useful~\cite{Clark-Londono-1991}.  Though the NonLens solutions were for a particularly constrained system, the third-order (`Seidel') aberration corrections can be applicable for cloaking designs. 

Another challenge is to minimize the visibility of unwanted rays, near the edges, in the paraxial designs. 
These edge effects become apparent in our cloaks and in unidirectional ray optics cloaks, 
when the incident angles are non-zero (with respect to the center axis of symmetry). 
The paraxial formalism does not correct for these, since edges are considered to be `infinitely' far away from the center-axis, and hence no longer `paraxial.'
So a paraxial cloak may be visible near the edges, although the paraxially valid region inside remains invisible for both the cloaked object and the cloaking device.
Though difficult to eliminate, the edge effects can be improved.
One method is to use the correct size for each lens or optical element, so as to match the optical design of the cloaks (See Fig.~\ref{fig:MultiClk2-4lens-sim}).  This was not necessarily the case for the cloaks demonstrated here.
Other strategies include reducing the total length of the system, using large optics, placing the cloaking device far away from observation, and using a thin outer shell to block stray light.  




Perhaps the most difficult design issue is increasing the incident angles, or field-of-view. 
The paraxial equations we presented are valid for angles $\theta$ that satisfy 
$\tan{\theta} \approx \sin{\theta} \approx \theta$.  
This is true for $\theta = 10^\circ - 15^\circ$, while even $\theta$ up to $30^\circ$ is considered to be valid for paraxial propagation~\cite{Siegman-book-1986}. 
Due to the choice of lenses, our cloaks were limited to several degrees.  However, we expect lens design techniques to be able to generate cloaks with much larger viewing angles, since ours relied on little, if any, such optimizations.  
To achieve larger field-of-view, balancing between lens radii of curvature and lens separation distances is needed.  However, this is not necessarily easy, particularly when image quality must also be maintained.  Optical engineering will be necessary to create cloaking devices that work for viewing angles beyond the paraxial regime, have high quality imaging, and allow for increased commercial utility.
Nonetheless, a small field-of-view can still be practical in cloaking satellites orbiting the earth, or for viewing distances that are far away. 

Lastly, the cloaking regions we designed require the center of the optical systems to not be obstructed. Also the cloaking region depended on the incident ray angles.
This was intentional so as to provide the simplest nontrivial solutions that demonstrated perfect paraxial cloaking for rays.
In the future, we will present methods that can cloak the center region (or a volume enclosed by light) instead, or allow cloaking regions that are independent of incident ray angles.




\section{Conclusion}
In summary, we have defined what a perfect cloak should do in ray optics.
We then provided a sufficient and necessary algebraic condition for a perfect cloak in the first-order, or paraxial, approximation. 
We finally derived a device that fits this definition, and experimentally demonstrated two cloaks for continuous ranges of directions.
In addition to hiding an object, these cloaking devices can also make an object behind a barrier visible, or deflect light rays.
Transformation optics and quasiconformal mapping are general formalisms used for cloaking fields. Here we provided another formalism that can effectively describe ray optics invisibility.


\appendix
\renewcommand\thesection{Appendix \Alph{section}}

\section{ABCD matrix background}
\label{app:ABCD-intro}
\begin{figure}[htbp]
\begin{centering}
\includegraphics[scale=1]{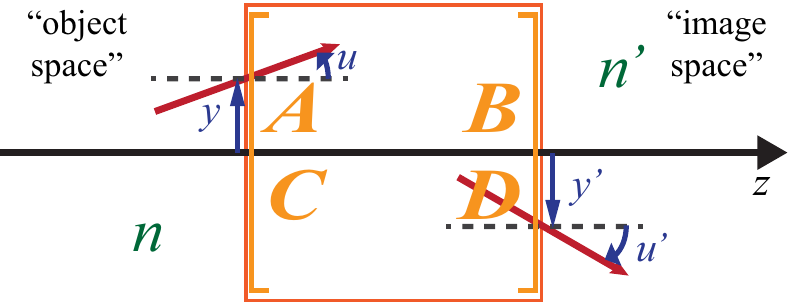}
\caption{\label{fig:ABCD-intro}
\textbf{Light rays and the `ABCD' matrix.}
Ray optics picture in the paraxial approximation.
We assume a rotationally symmetric system (about the $z$-axis), with light traveling from left to right. 
The optical system (box in the center) can be described by an `ABCD' matrix. This matrix maps the initial position ($y$) and paraxial angle ($u$) to those exiting the system ($y'$, $u'$). The ``object space'' is the space before the ABCD system, with index of refraction $n$. Likewise, the ``image space'' is the space after the system, with index of refraction $n'$.
In this diagram, $y>0$, $u>0$, $y'<0$, and $u'<0$, for our sign convention~\cite{Buralli-Geo-1}.}
\end{centering}
\end{figure}

We borrow well-known linear equations from the field of geometric optics~\cite{Handbook-Opt-v1-2010}. 
We will assume rotational symmetry for the system. 
To first approximation, called the ``paraxial approximation,'' light rays are assumed to deviate minimally from the axis of rotational symmetry for the system ($z$ in Fig.~\ref{fig:ABCD-intro}). We can then see that in the paraxial approximation, ray angles are small.

All light rays can then be described by its transverse position $y$ and the paraxial angle $u$, much like a particle can be described by its position and momentum.  
Here, 
\be
u \equiv 
\tan{\theta}
\approx \theta
\label{eq:u}
,
\ee
where $\theta$ is the exact, real angle of the ray from the $z$-axis.

Because of the linearity of optics in the paraxial approximation, the propagation of light rays through an optical system can be described by matrices. 
Utilization of matrices for ray optics has a rich history, with equations that originated from Newton and Gauss~\cite{Nussbaum-book-1998,Nussbaum-1978}.
These matrices are called `ray transfer' matrices, or `ABCD' matrices, and are used as follows:
\be  
\begin{bmatrix}
y' \\
n' u'
\end{bmatrix}
=
\begin{bmatrix}
A & B \\
C & D
\end{bmatrix}
\begin{bmatrix}
y \\
n u
\end{bmatrix}
,
\label{eq:ABCD-define}
\ee
where $n$ is the index of refraction, $y$ is the   transverse position, and $u$ is the paraxial angle, immediately before the ABCD matrix. $n', y'$, and $u'$ are those for after the matrix (See Fig.~\ref{fig:ABCD-intro}).
For example, the ABCD matrix for a space of length $t$, with index of refraction $n_t$, is the `translation matrix' (or `transfer matrix'):
\be  
M_t =
\begin{bmatrix}
1 & t/n_t \\
0 & 1
\end{bmatrix}
.
\label{eq:translation-matrix}
\ee

\section{ABCD matrix for a cloak}
\label{app:ABCD-cloak}
So what does the ABCD matrix for a perfect cloaking system look like? It is precisely the translation matrix $M_t$ in \myeq{translation-matrix}, where $t=L$ is the length of the system, and $n_t=n=n'$. This is because a perfect cloaking device simply replicates the surrounding medium throughout its volume. 
We then see that
\be  
\begin{bmatrix}
y' \\
n u'
\end{bmatrix}
=
\begin{bmatrix}
1 & L/n \\
0 & 1
\end{bmatrix}
\begin{bmatrix}
y \\
n u
\end{bmatrix}
=
\begin{bmatrix}
y + L u \\
n u
\end{bmatrix}
.
\ee
As expected, the angle remains the same, i.e. $u'=u$, and the position shifts by the angle multiplied by the length, i.e. $y'=y + L u$.

So far we assumed the ambient medium of the cloaking device to be uniform. Suppose this was not the case. 
In the paraxial approximation, we can then assume the changes happen along $z$ only, since transverse deviations are small. 
Thus we can write $n\rightarrow n(z)$ to account for the nonuniform medium. 
Let's assume there are $N$ segments along $z$ within the device, such that $n(z) \equiv n_i$ is constant for the region $z \in [z_{i-1}, z_{i})$.
Here, $z=z_0$ at the beginning of the device, and $z=z_N=(z_0+L)$ at the end of the device (of length $L$).
Then, the ABCD matrix for a perfect cloak can be written as
\bea
&&
\prod_{i=1}^N
\begin{bmatrix}
1 & (z_i-z_{i-1})/n_i \\
0 & 1
\end{bmatrix}
= 
\label{eq:cloak-metric-nonuniform}
\\
&& \!
\begin{bmatrix}
1 & (z_N \!-\! z_{N-1})/n_N \\
0 & 1
\end{bmatrix}
\! \cdots \!
\begin{bmatrix}
1 & (z_2 \!-\! z_{1})/n_2 \\
0 & 1
\end{bmatrix}
\! \cdot \!
\begin{bmatrix}
1 & (z_1 \!-\! z_0)/n_1 \\
0 & 1
\end{bmatrix}
\!.
\nonumber
\eea

\section{Methods}
\label{app:methods}
It is important to point out that the ABCD matrix equations above assumed that all the lenses are infinitely thin. In reality, lenses have thicknesses, and this can make a difference in obtaining the correct solutions for 
\myeq{cloak-metric}. 
So to set up our simulations, we took a step further to include the lens thicknesses and materials.

\subsection{Three lens symmetric cloak}
For the first and last lenses (1 and 3) we used 200 mm focal length, BK7, 75 mm diameter lenses.
For the center lens, we used two -100 mm focal length, BK7, 50 mm diameter lenses, back-to-back, to create a lens with focal length $\approx\!\!-50$ mm.
All lenses were catalogue optics from Edmund Optics.
From 
\myeq{3lens-f2C-symm}, 
we obtain $t_1=t_2\approx 100$ mm. 
Including the lens thicknesses and the material indices of refraction, we optimized $t_1$ slightly so that the afocal condition $C=0$ was closely achieved.
Diameter of last lens needs to be $>150$ mm for all rays to pass (no ``vignetting'').
For the CODE V simulation, the apertures were not restricted to the actual lens sizes.
The aperture stop was the first surface.
Aperture diameter sizes (for no vignetting) of the first and second diverging lenses in the center are 54 mm and 61 mm, respectively.
Total length of the system is 219 mm.

Proper alignment is important for the cloaking effect.  We analyzed the sensitivity of the system due to perhaps the easiest misalignment to occur- changes in the lens separation, i.e. $t_1$ and $t_2$ for the three lens case.
As stated previously, $t_1=t_2$ with the ideal value being $\approx 100$ mm.
With a $\pm 1\%$ change in $t_1$, the magnitude of the effective focal length (for infinite conjugates) decreased by a factor of $\approx 40$, and the magnification changed by $\approx 10\%$ from $1.0$.
For a $\pm 10\%$ change in $t_1$, the magnitude of the effective focal length (for infinite conjugates) decreased by a factor of $\approx 400$, and the magnification changed by a factor of $\approx$ -20 and 0.5.

Using CODE V, aberrations were calculated for the simulation given in Fig.~\ref{fig:MultiClk2-3lens}(a), but for an object distance of 2 m from the first surface, and an image distance of 5.3 m from the last surface.
For the nominal wavelength (587.6 nm), the total surface contributions (in mm) for the third-order 'structural' aberrations
($\sigma$ coefficients for the transverse ray errors~\cite{Buralli-Geo-1,Bentley-Opt444-2012}) 
were 4.3, -12.4, 8.4, -1.0, -4.3
for spherical aberration, coma, astigmatism, Petzval, and distortion, respectively.  
The CODE V generated axial and lateral colors were 7.7 mm and -2.9 mm, respectively, with a total distortion of $1.4\%$ for the $-3.5^\circ$ field angle.  
Most of the aberrations occurred at the last (third) lens.  These aberrations are expected, since only $t_1$ was numerically optimized (and $t_2$ matched to $t_1$) to produce a nearly afocal system, while no aberration corrections were made.

\subsection{Four lens symmetric cloak}
For the first and last lenses (1 and 4), we used 200 mm focal length, 50 mm diameter achromatic doublets composed of BK7 and SF2 glasses. For the center two lenses (2 and 3), we used 75 mm focal length, 50 mm diameter achromatic doublets composed of SF11 and BAF11 glasses. 
All doublets were off-the-shelf catalogue optics from Thorlabs and had anti-reflection coating in the visible spectrum.
For the CODE V simulations, the aperture sizes were not restricted, so as to ensure no vignetting.
The aperture stop was the first surface.
Diameters of the second, third, and last doublets need to be $>$ 33 mm, 51 mm, and 112 mm, respectively, for no vignetting.
Total length of the system is 910 mm.

Again, we analyzed the sensitivity of the system due to lens separation misalignments, i.e. $t_1, t_2$, and $t_3$ for the four lens cloak.
Since ours is a symmetric cloak, $t_1=t_3$, which was maintained.
Only one variable was changed at a time, either $t_1=t_3$ or $t_2$, and not both, for this analysis.
With a $\pm 1\%$ change in $t_1$, the magnitude of the effective focal length (for infinite conjugates) decreased by a factor of $\approx 5\times10^4 \sim 6\times10^4$, and the magnification changed by $\approx 30-40\%$ from $1.0$.
For a $\pm 10\%$ change in $t_1$, the magnitude of the effective focal length (for infinite conjugates) decreased by a factor of $\approx 3\times10^5 \sim 8\times10^5$, and the magnification changed by a factor of $\approx$ -0.8 and 0.2.
No changes in these values were seen for $1\%\sim10\%$ changes in $t_2$, since $t_2$ does not affect the on-axis fields at 0 degrees.

Aberrations were calculated for the simulation given in Fig.~\ref{fig:MultiClk2-4lens-sim}(b) using CODE V, but for an object distance of 2 m from the first surface, and an image distance of 3 m from the last surface.  For the nominal wavelength (587.6 nm), the total surface contributions (in mm) for the third-order 'structural' aberrations
($\sigma$ coefficients for the transverse ray errors~\cite{Buralli-Geo-1,Bentley-Opt444-2012}) 
were 10.4, -8.6, 8.2, 0.4, -8.1
for spherical aberration, coma, astigmatism, Petzval, and distortion, respectively.  
The CODE V generated axial and lateral colors were 2.4 mm and -1.3 mm, respectively, with a total distortion of $7.7\%$ for the $-1.5^\circ$ field angle.  
Most of the aberrations occurred at the last two achromats.  These aberrations are not surprising, given that no optimization was performed, other than the use of achromats.

\section*{Acknowledgments}
This work was supported by the Army Research Office Grant No. W911 NF-12-1-0263 and the DARPA DSO Grant No. W31P4Q-12-1-0015.
The authors would like to acknowledge the helpful geometric optics discussions with Aaron Bauer, Robert Gray, and Kyle Fuerschbach, as well as ideas shared by Greg Howland. 


\begin{thebibliography}{10}
\newcommand{\enquote}[1]{``#1''}

\bibitem{Gbur-Cloak-2013}
G.~Gbur, \enquote{Invisibility physics: Past, present, and future,} Prog.\ Optics \textbf{58}, 65--114 (2013).

\bibitem{Leonhardt-2006}
U.~Leonhardt, \enquote{Optical conformal mapping,} Science \textbf{312},
  1777--1780 (2006).

\bibitem{Pendry-2006}
J.~B. Pendry, D.~Schurig, and D.~R. Smith, \enquote{Controlling electromagnetic
  fields,} Science \textbf{312}, 1780--1782 (2006).

\bibitem{McCall-2013}
M.~McCall, \enquote{Transformation optics and cloaking,} Contemp.\ Phys.\
  \textbf{54}, 273--286 (2013).

\bibitem{Zhang-review-2012}
B.~Zhang, \enquote{Electrodynamics of transformation-based invisibility
  cloaking,} Light.\ Sci.\ Appl.\ \textbf{1}, e32 (2012).

\bibitem{Schurig-2006}
D.~Schurig, J.~J. Mock, B.~J. Justice, S.~A. Cummer, J.~B. Pendry, A.~F. Starr,
  and D.~R. Smith, \enquote{Metamaterial electromagnetic cloak at microwave
  frequencies,} Science \textbf{314}, 977--980 (2006).

\bibitem{Landy-Smith-2013}
N.~Landy and D.~R. Smith, \enquote{A full-parameter unidirectional metamaterial
  cloak for microwaves,} Nat.\ Mater.\ \textbf{12}, 25--28 (2013).

\bibitem{Li-Pendry-2008}
J.~S. Li and J.~B. Pendry, \enquote{Hiding under the carpet: A new strategy for
  cloaking,} \prl \textbf{101}, 203901 (2008).

\bibitem{Ergin-2010}
T.~Ergin, N.~Stenger, P.~Brenner, J.~B. Pendry, and M.~Wegener,
  \enquote{Three-dimensional invisibility cloak at optical wavelengths,}
  Science \textbf{328}, 337--339 (2010).

\bibitem{Soric-Alu-Mantle-2013}
J.~C. Soric, P.~Y. Chen, A.~Kerkhoff, D.~Rainwater, K.~Melin, and A.~Alu,
  \enquote{Demonstration of an ultralow profile cloak for scattering
  suppression of a finite-length rod in free space,} New J.\ Phys.\
  \textbf{15}, 033037 (2013).

\bibitem{Rainwater-Alu-3D-2012}
D.~Rainwater, A.~Kerkhoff, K.~Melin, J.~C. Soric, G.~Moreno, and A.~Alu,
  \enquote{Experimental verification of three-dimensional plasmonic cloaking in
  free-space,} New J.\ Phys.\ \textbf{14}, 013054 (2012).

\bibitem{Fridman-2012}
M.~Fridman, A.~Farsi, Y.~Okawachi, and A.~L. Gaeta, \enquote{Demonstration of
  temporal cloaking,} Nature \textbf{481}, 62--65 (2012).

\bibitem{Lukens-2013}
J.~M. Lukens, D.~E. Leaird, and A.~M. Weiner, \enquote{A temporal cloak at
  telecommunication data rate,} Nature \textbf{498}, 205--208 (2013).

\bibitem{Brule-2014}
S.~Brule, E.~H. Javelaud, S.~Enoch, and S.~Guenneau, \enquote{Experiments on
  seismic metamaterials: Molding surface waves,} \prl
  \textbf{112}, 133901 (2014).

\bibitem{Zhang-2011}
B.~L. Zhang, Y.~A. Luo, X.~G. Liu, and G.~Barbastathis, \enquote{Macroscopic
  invisibility cloak for visible light,} \prl \textbf{106},
  033901 (2011).

\bibitem{Chen-2011}
X.~Z. Chen, Y.~Luo, J.~J. Zhang, K.~Jiang, J.~B. Pendry, and S.~A. Zhang,
  \enquote{Macroscopic invisibility cloaking of visible light,} Nat.\ Comm.\  \textbf{2}, 176 (2011).

\bibitem{Chen-2013}
H.~Chen, B.~Zheng, L.~Shen, H.~Wang, X.~Zhang, N.~I. Zheludev, and B.~Zhang,
  \enquote{Ray-optics cloaking devices for large objects in incoherent natural
  light,} Nat.\ Comm.\ \textbf{4}, 2652 (2013).

\bibitem{Zhai-2013}
T.~R. Zhai, X.~B. Ren, R.~K. Zhao, J.~Zhou, and D.~H. Liu, \enquote{An
  effective broadband optical 'cloak' without metamaterials,} Laser Phys.\ Lett.\ \textbf{10}, 066002 (2013).

\bibitem{Howell-Cloak-2014}
J.~C. Howell, J.~B. Howell, and J.~S. Choi, \enquote{Amplitude-only, passive,
  broadband, optical spatial cloaking of very large objects,} \ao
  \textbf{53}, 1958--1963 (2014).

\bibitem{Wolf-Cloak-1993}
E.~Wolf and T.~Habashy, \enquote{Invisible bodies and uniqueness of the inverse
  scattering problem,} \jmo \textbf{40}, 785--792 (1993).

\bibitem{Nachman-1988}
A.~I. Nachman, \enquote{Reconstructions from boundary measurements,} Ann.\ Math.\ \textbf{128}, 531--576 (1988).

\bibitem{Devaney-1978}
A.~J. Devaney, \enquote{Nonuniqueness in inverse scattering problem,} J.\ Math.\ Phys.\ \textbf{19}, 1526--1531 (1978).
  
\bibitem{Born-Wolf-2010}
M.~Born and E.~Wolf, \emph{Principles of Optics: Electromagnetic Theory of
  Propagation, Interference and Diffraction of Light} (Cambridge University, 2010), 7th ed.

\bibitem{Siegman-book-1986}
A.~E. Siegman, \emph{Lasers} (University Science Books, 1986).

\bibitem{Handbook-Opt-v1-2010}
M.~Bass, \emph{Handbook of Optics- Geometrical and Physical Optics, Polarized Light, Components and Instruments} (McGraw-Hill, 2010), \textrm{Vol.}~1, 3rd ed.

\bibitem{FG-GO-2004}
J.~E. Greivenkamp, \emph{Field Guide to Geometrical Optics} (SPIE, 2004).

\bibitem{Clark-Londono-1991}
P.~P. Clark and C.~Londono, \enquote{1990 International Lens Design Conference
  lens design problems: the design of a NonLens,} in \enquote{1990 Intl Lens
  Design Conf,} (Proc. SPIE, 1991), \textbf{1354}, pp. 555--569.

\bibitem{Buralli-Geo-1}
D.~Buralli, \emph{OPT 441- Geometrical Optics} (The Institute of Optics,
  University of Rochester, 2010).

\bibitem{Nussbaum-book-1998}
A.~Nussbaum, \emph{Optical System Design} (Prentice Hall PTR, 1998).

\bibitem{Nussbaum-1978}
A.~Nussbaum, \enquote{Teaching of advanced geometric optics,} \ao
  \textbf{17}, 2128--2129 (1978).

\bibitem{Bentley-Opt444-2012}
J.~Bentley, \enquote{Optics 444- Lens Design,}  (2012). Course lecture notes.

\end{thebibliography}
\end{document}